Title: The Impact of Coupling on Estimation of Local and Global SAR with Transmit Arrays

Pei-Shan Wei,[1,2] Christopher P. Bidinosti,[1,3] Scott B. King[1,4]

[1]Department of Physics & Astronomy, University of Manitoba, Winnipeg, Canada R3T2N2

[2]Sunnybrook Research Institute, Physical Sciences Platform, Toronto, ON, Canada M4N3M5

[3]Department of Physics, University of Winnipeg, Winnipeg, Canada R3B2E9

[4] Philips Healthcare - Invivo, Gainesville, FL, USA 32608

# Abstract

In pTx MRI systems, the prediction of local SAR is based on numerical electromagnetic (EM) simulations and used to scale RF power to ensure FDA SAR limits are not exceeded. This prediction becomes more complex when superposition of E-fields from multiple coupled coils are employed in parallel transmission, each affected by dielectric and conductive properties of the human body. It was demonstrated that incorrect inductive coupling used in simulations of transmit array coil spatial excitation and SAR, leads to poor accuracy of predicted excitation and SAR, and more importantly from a safety perspective, underestimated local SAR by 19-40%.

## 1. INTRODUCTION

Parallel transmit (pTx) MRI is a state-of-the-art technique which has many applications, most notably [1] RF shimming to compensate for wave behavior leading to so called central brightening [1], [2]; and [2] spatially-selective RF excitation [3]–[5] which can be used for small field-of-view (FOV) excitation [6], [7] or compensation for $B_0$ inhomogeneity by incorporating a $B_0$ map into RF pulse design [8]. Parallel transmit technology is particularly important in high-field (≥ 3 Tesla) functional MRI studies and cardiac MRI [9]–[11] where $B_1$ and $B_0$ field inhomogeneity can drastically hinder imaging performance. Small FOV RF excitation could improve cardiac MRI tremendously by enabling higher spatial or temporal resolution [12]. However, when applying small FOV RF excitation with single-channel transmission, unsuitably long multi-dimensional RF pulses are required. Transmit SENSE (Tx-SENSE) [7], [13], [14] on the other hand allows much shorter multi-dimensional pulses. A challenge with this method is that it relies on the superposition of multiple, predetermined RF waveforms, and incorrect RF amplitude and phase applied to the multi-channel transmit array may lead to unknown local SAR hot spots. As a result, SAR predictions must be considered carefully here, and a conservative strategy is best.

As $B_0$ magnitude, and hence $B_1$ frequency, increases, the RF power dissipated as heat in the human body becomes a major concern for safety. This is a major concern with high-field systems, given that the RF power dissipation increases roughly with the square of the $B_0$ field strength [15]. In pTx MRI systems, the prediction of local SAR is based on numerical electromagnetic (EM) simulations. During a scan, the actual RF power applied during a specific pulse sequence is rescaled based on these simulated

SAR values and is limited by FDA guidelines [16]. However, when multiple transmit channels are employed in parallel transmission, the prediction becomes more complex. The induced E-field distribution is a superposition of individual E-field distributions for each channel and may cause a high local SAR (i.e., a hot spot) due to constructive interference of the individual E-field distributions. In addition, the E-field distributions for individual transmit channels are affected by dielectric and conductive properties of the human body [17], causing further asymmetries in the induced E-field distributions.

Although it is possible to monitor global SAR during pTx MRI [18], there is no real-time in-vivo method of monitoring local SAR or temperature changes. Further complications arise in pTx MRI because complex time-varying amplitude and phase modulations of the RF waveforms are applied to each transmit channel to form irregular spatial excitations, which in turn leads to local SAR hot spot distributions that are difficult to predict.

The performance of Tx-SENSE depends on RF coil design and RF pulse design, together with an accurate prediction of local SAR distribution. Clearly, design goals that include accurate predictions of local SAR and minimization of local SAR hot spots are particularly important for patient safety in parallel transmission MRI [19]. The RF coil array for pTx systems determines both the $B_1^+$ maps used for Tx-SENSE RF pulse design and the simulations of induced electric field distributions for SAR estimation. It is well known that inductive and resistive coupling between array elements affects the resulting $B_1^+$ maps and E-field distributions [19]. Therefore, to optimize transmit array (Tx-array) design and predict local SAR distribution through simulations, it is important to take into account the impact of array coupling.

A question remains: do the predictions resulting from the simulated local SAR distribution accurately represent the actual experimental situation, where both array layout and coupling may vary from the simulation model? Coil elements are coupled to each other, especially during transmission. Typically, this coupling is not taken into account in a simulation. However, the problem is that this coupling will change the field distributions, and the actual Tx-array may not be identical to the simulated array in terms of coupling. Hence, our goal is to investigate how inductive coupling differences in transmit array coil design/constructions influence the design of the RF pulse, as well as the accuracy of predicted excitation and SAR. A Tx-array coil was built and subsequently used to validate that coupling can indeed be introduced into simulations.

## 2. METHODS

The general procedure of optimizing coil design for Tx-SENSE is shown in Fig. 1. Here coil design starts with EM simulations of individual Tx-channels to give $B_1^+$ maps and E-field distributions, which are the outputs. The $B_1^+$ maps are used in the design of Tx-SENSE RF pulses for individual channels. The waveforms of these RF pulses are then used as weighting factors of individual E-field distributions for SAR estimation. In addition, the $B_1^+$ maps are used along with the RF pulses to simulate the small FOV excitation. Using the constructed coil, we acquired $B_1^+$ maps in a MR scanner to compare with the simulated ones.

## 2.1. RF Coil Configuration

To evaluate the impact of coupling, an eight-channel transmit-array coil was modeled in a finite difference time-domain (FDTD)-based electromagnetic simulator (SEMCAD X v14.8, SPEAG, Zurich, Switzerland). This coil was composed of anterior and posterior sections (Fig. 2) as five and three overlapping coil elements respectively. The overall dimensions of these sections were 22.7 cm wide by 19 cm in length. Each coil element contained either seven or six capacitors (28 or 20 pF) and was tuned to 123.2 MHz. All simulations were performed with a cardiac torso model (60 cm x 47.4 cm x 24 cm) shown in Figure 2. To address the impact of inductive coupling, EM fields were calculated for the three scenarios shown in Fig. 3 based on the model in Fig. 2. Here, "isolation" (Fig. 3A) is where each element was simulated separately without including any inductive coupling, and "overlap decoupled" (Fig. 3B) is with all elements active and represents how the actual coil was constructed. In this case, it is important to note that the next neighbor coil elements are still coupled. When a coil is constructed, the layout of the finished coil may be slightly different from the original design. To mimic this, a coil element was slightly shifted from its optimal position for overlap decoupling as illustrated in Fig. 3C. Here, "coupled" is where each element was simulated in the presence of other coupled elements that were terminated with 50 Ω.

To validate the impact of coupling on the $B_1^+$ maps, the coil based on the posterior section in Fig. 2 was constructed and is shown in Fig. 4, where the longitudinal conductors of the central element (#1) were adjusted for zero mutual inductance with the neighboring elements. For simplicity, the anterior section was not used for validation. The coil was made from a copper clad printed circuit board that was milled to give the layout of the three coil elements with a copper trace width of 5 mm. Each coil element was tuned and matched to 123.2 MHz using a calibrated network analyzer (Agilent E5601A). A preamplifier (MPB-123R20-90, Hi-Q.A. Inc., Carleton Place, ON, Canada), cable trap, T/R switch and matching circuit were added to each coil element. The double-angle method [20] was used to acquire the $B_1^+$ maps in a 3T MR scanner (Siemens, Trio). The parameters used in this gradient echo sequence were the following: acquisition matrix size = 128 x 512, TR/TE = 2000/15 ms and slice thickness = 5 mm. To mimic an isolated coil, i.e. Fig. 3A, only the central element was used to transmit and receive the MR signal; the other elements were physically open. The coupled scenario illustrated in Fig. 3C was also implemented on the constructed coil by shifting one leg of the central element by 3mm. The two outer elements (#2, #3) were not geometrically decoupled from each other. They were resonant loops and were receive-only.

## 2.2. RF Pulse Design for Tx-SENSE

To evaluate the impact of coupling on the multi-Tx array system, two sets of RF pulses were designed; one is based on physically isolated coil elements and the other on more realistic overlap decoupled coil elements. The custom RF pulse design used here was a spatial-domain based method and it determined a set of RF pulses by finding a pseudo-inversion of a least-squares problem [6]. These RF pulses used as input, $B_1^+$ maps generated from either isolation or coupled simulations using the eight-channel Tx-array shown in Fig. 2. The excitation pattern was a homogenous box covering the cardiac region of the phantom. The parameters used for RF pulse calculation were the following: FOV = 55.5 cm x 35.8 cm, excitation FOV = 12 cm x 12 cm, object matrix size = 140 x 91, maximum amplitude of gradient = 40 mT/m, slew rate of the gradient = 200 T/m/s, spiral excitation k-space, number of spiral turns ($N_t$) = 16, and the length of RF pulse ($T_p$) = 5.1, 3.9, 3.1 and 2.6 ms associated with α = 1, 2, 3, and 4 in variable density k-space trajectory [21]. The α term determines the amount of oversampling near the origin of the k-space trajectory.

## 2.3. SAR Estimation in Tx-SENSE

In order to estimate SAR for Tx-SENSE, the value of SAR per pixel was calculated using

$$SAR(\vec{r}) = \frac{\sigma(\vec{r})}{2\rho(\vec{r})T} \int_0^T \|E(\vec{r},t)\|^2 dt$$

(1)

where $E(\vec{r},t) = \sum_{p=1}^{P} a_p(t) E_p(\vec{r})$

$\sigma$ and $\rho$ are the conductivity and density of the phantom, $\vec{r}$ is the spatial position of each voxel, and $T$ is the pulse length. Scalar weighting factors $a_p$ are the amplitudes of RF waveforms at time points $t$. The total E-field distribution is a spatial-temporal function and is a superimposition of individual E-fields ($E_p(\vec{r})$) of the $P$ transmit channels. The applied RF pulses have complicated unknown waveforms in small FOV spiral excitation. Therefore, when we superimpose all the E-field distributions, a local hot spot can form at a location that is difficult to predict

As the parameters for RF pulse design are changed (e.g. k-space trajectory), the weighting factors for SAR must be associated with the new set of RF pulses and change correspondingly. Local SAR is calculated per voxel and then the maximum value of local SAR is found, which is commonly called a hot spot or peak SAR. Global SAR is also calculated by averaging the SAR values over the whole volume of the phantom, and is used as a scaling factor for the total power transmitted by the RF coil. For a given RF coil design, local and global SAR will vary with individual transmit RF pulses and is therefore a criterion used in optimizing the RF coil design.

## 3. RESULTS

To validate the impact of inductive coupling, the posterior section of the eight-channel Tx-array was simulated and the EM fields determined. Fig. 5A shows the resulting $B_1^+$ map of the central coil element. Enlargements of the dashed region show differences between the isolated scenario (Fig. 5C) and the coupled scenario (Fig. 5E). For validation, $B_1^+$ maps were acquired in the MR scanner using the constructed coil with an identical configuration as the simulated coil. We acquired $B_1^+$ maps for the isolated scenario (Fig. 5B and 5D) and the coupled scenario (Fig. 5F). The measured $B_1^+$ maps show good qualitative agreement with the simulated $B_1^+$ maps. It can be seen that all $B_1^+$ maps are asymmetrical, but the coupled $B_1^+$ maps have fields extending towards the left-hand side (shown by the arrow) where the coupled neighboring element was located. This is expected because the cross talk between coil elements would share the field distribution.

If RF pulses are determined using incorrect $B_1^+$ and E-field maps, the predictions of the excitation profile and SAR could be wrong. To highlight the influence of the three coupling scenarios on the resulting excitation profiles, Fig. 6 shows Bloch simulations with four different combinations of RF pulses and $B_1^+$ maps. Assuming ideal uncoupled fields existed (Fig. 3A) in a MR scanner, results in the excitation shown in Fig. 6A. This is the result of RF pulses determined from isolated fields (Fig. 3A), i.e. no coupling, Bloch simulated with isolated fields (Fig. 3A). It can be seen that there are excitation artifacts near the top center of the image close to the position of the coil element. Because coupling between next neighboring coil elements always exists, we applied RF pulses determined from coupled fields (Fig. 3B) to a Bloch simulation using coupled fields (Fig. 3B), where fields were generated in the overlap decoupled scenario where only the neighboring coil elements were overlap-decoupled (Fig. 3B). The resulting excitation profile is shown in Fig. 6B and this produced less excitation artifacts close to the surface of the phantom as compared to the result for the isolated scenario shown in Fig. 6A.

Within the procedure for small FOV excitation, one may try to apply RF pulses determined from isolated fields (Fig. 3A) to an experimental Tx-array that includes coupling (Fig. 3B). To mimic this procedure, "experimental" coupled fields were simulated by including crosstalk between the simulated coil elements as shown in Fig. 3B. The Bloch simulation excitation profile resulting from this procedure is shown in Fig. 6C. It can be seen that the average flip angle inside the target region is lower than the target flip angle.

Additionally, one might also consider an experimental Tx-array that may not have perfect overlap decoupling (Fig. 3C), but RF pulses are determined from an assumed perfect overlap decoupled fields (Fig. 3B). This is shown in Figure 6D. In this case, there are more excitation artifacts and the target pattern is less uniform. This excitation profile has a poorer performance than for a properly overlap decoupled coil as shown in Fig. 6B.

For the small FOV excitation described above, we found that coupling had significant effect on $B_1^+$ maps and the corresponding determination of RF excitation pulses. Subsequently, SAR distributions were expected to be directly affected by the combined effect of RF pulses and E-field distributions as described by Equation 1. To characterize this effect, we calculated SAR values for each voxel of the phantom for the same four coupling combinations used for Fig. 6. The results for the isolated scenario, i.e. no coupling, are shown in Fig. 7A. When the effect of coupling is taken into account, Fig. 7B shows that the prediction of the location of the hot spots shown in Fig. 7A could be wrong.

Here we can see the hot spots appear at the anterior region for the isolated scenario (Fig. 7A), but the hot spots move to the posterior region in the overlap decoupled scenario (Fig. 7B).

The SAR map shown in Fig. 7C is generated for the case of applying RF pulses determined from isolated fields (Fig. 3A) to the case where the E-field distributions were determined from an overlap decoupled scenario (Fig. 3B). In this case, the RF power was not scaled and would need to be increased to achieve the designed target excitation pattern. To demonstrate the potential error in predicting hot spots resulting from variations in coil construction, Fig. 7D shows a SAR map generated by the RF pulses determined for the overlap decoupled scenario (Fig. 3B) but applied to E-field distributions in a coupled coil (Fig. 3C). It can be seen that this SAR distribution has small local SAR variations as compared to Fig. 7B which uses overlap decoupled fields for both designing the RF pulses and for calculating SAR values.

Figures 6 and 7 show how RF pulses determined from different coupling scenarios affect both Tx-SENSE performance and SAR distributions. It is also known from Equation 1 that SAR is also predicted to depend on pulse length. To quantify this effect, the value of SAR per voxel over the phantom was calculated and used to determine global SAR, peak SAR, and the ratio of peak SAR to global SAR. These were calculated as a function of RF pulse length and are shown in Fig. 8 and Fig. 9 for various coupling combinations. Fig. 8 illustrates when both RF pulses and the EM fields were determined in either isolation (scenario 1) or with overlap decoupling (scenario 2). It can be seen that the length of the RF pulse ($T_p$) significantly altered both global and peak SAR estimation. Both coupling scenarios had minimum global SAR and peak SAR values at $T_p$ = 3.9 ms.

Notably, global SAR was overestimated by up to 23% at $T_p$ = 2.6 ms when the incorrect coupling scenario (isolated) was used, whereas peak SAR using the incorrect coupling scenario was underestimated by 21% at $T_p$ = 3.9 ms to as much as 39% at $T_p$ = 3.1 ms. Similarly, the ratio of peak SAR to global SAR for the incorrect coupling scenario (isolated) was always underestimated with an error that varied with $T_p$. Specifically, the errors are 33,%, 40%, 31%, and 20%,  for $T_p$ values of 2.6, 3.1, 3.9, and 5.1 ms respectively.

In practice, determining the parameters of RF pulses using the overlap decoupled scenario is more accurate than assuming the inductive coupling is zero. However, if a constructed coil generates EM fields slightly different than those predicted by simulations, both small FOV excitation and SAR maps would be adversely altered (see Fig.6 and Fig. 7). Hence, to estimate how the SAR evaluation was affected by the variations between the constructed and simulated coils, the coil configuration shown in Fig. 3C was used. Recall that this configuration had one coil element that was shifted by 3 mm and thus the overlap decoupling was not perfect. Coupled E-field distributions (from this coil) along with RF pulses determined in the (perfect) overlap decoupled scenario were used to determine SAR values. These are plotted as a function of the RF pulse length in Fig. 9. For reference, we also show SAR values for the perfect case resulting from overlap decoupled E-field distributions and overlap decoupled RF pulses. The minimum global SAR and peak SAR are at $T_p$ = 3.9 ms for both scenarios. The worst case predictions for SAR values are: the global SAR is overestimated by 26% at 2.6ms, 36% at $T_p$ = 3.1 ms; peak SAR is underestimated by 20% at $T_p$ = 2.6 ms and < 10% at other $T_p$; and the ratio of peak SAR to global SAR is underestimated by 37%, 30%, 31%, and 19% at $T_p$ = 2.6, 3.1, 3.9, 5.1 ms respectively.

## 4. DISCUSSION

We have investigated how different coupling scenarios affected small FOV excitation and SAR estimation. Our goal was to quantify the impact of coupling and to determine the degree of under- or over-estimation of local and global SAR. If the designed RF pulses and the $B_1^+$ maps are mismatched in the Bloch simulation of small FOV excitation, the resulting excitation profiles will be altered and this will lead to a wrong interpretation of SAR estimation.

For validation, simulations of $B_1^+$ maps were compared with experimental measurements from the posterior section of a constructed eight-channel transmit coil array shown in Fig. 5. The input power of the simulation was fixed at 1 Watt, but the transmit power of the MR experimental B1 mapping was much higher to achieve the flip angles required for the B1 mapping technique. Therefore, the amplitudes of the simulated and measured $B_1^+$ maps cannot be directly compared. However, the qualitative agreement between the simulations and measurements suggests that the simulated field is able to appropriately incorporate the coupling information. Therefore, the simulations were used for further analysis of coil design.

The E-field distribution of the coil array cannot be directly measured on a scanner, but we can compare experimental $B_1^+$ maps for each transmit channel to simulation, and if there is good agreement, than we can infer that E-field simulations could also be used as a predictor of experimental SAR in various coupling conditions. The constructed coil was slightly different than the simulated coil model because the model was simplified by ignoring some electric components, such as the detuning circuitry. Thus, the simulated E-fields and the experimental E-fields were not expected to be identical. However, Fig.

7D shows that even though small variations in coil construction are likely, predictions of the general locations of hot spots assuming an overlap decoupled scenario remain valid.

Simulations of small FOV excitation and SAR estimation showed that predicted results were affected, to different degrees, by how the RF pulses were designed. Simulation of small FOV excitation required $B_1^+$ maps, i.e. sensitivity maps, for each transmit channel. It also required the application of RF pulses that were calculated using these same $B_1^+$ maps. Additionally, the RF pulses needed to take into account a suitable k-space trajectory to control the image resolution and the pulse length. Hence, the design of an RF pulse was affected by RF pulse length and coupling and these impacted the performance of Tx-SENSE. The $B_1^+$ maps required for Bloch simulations were generated using the different coupling scenarios illustrated in Fig. 3. These coupling scenarios are the ones most commonly considered during RF coil simulation and construction.

For the estimation of local SAR, it is very important to consider coupling while designing the RF pulses and determining the associated EM fields. This is illustrated in Fig. 6C where the average flip angle inside the target region was lower than the target flip angle. Therefore, the RF power for this combination of RF pulse and $B_1^+$ maps must be scaled upwards to achieve the designed target pattern. In addition, the absolute intensities of the hot spots in Fig. 7C seem lower than the other cases, but this is an unscaled SAR distribution and would lead to an underestimation of the local SAR value. Hence, once coupling has been included, scaling of RF power is vital for accurate SAR estimation.

We have shown the necessity of incorporating inductive coupling when simulating Tx-arrays used for Tx-SENSE. If the effect of coupling is ignored, the prediction of hot spots can be wrong. We also found SAR values varied with various lengths of RF pulses as demonstrated in Fig. 8. These results show that with incorrect coupling, global SAR was overestimated by as much as 23%, peak SAR underestimated by as much as 39%, and the ratio of peak-to-global SAR underestimated by as much as 40. For any pulse length, the percentage difference of SAR values between the isolation and overlap decoupled scenarios was significant. This percentage difference is remarkably similar to published findings for a whole-body TEM Tx-array coil where the SAR values were over or under estimated by 20~ 40% [22]. For the case where the constructed coil array was slightly different from the simulation model of a coupled coil array (see Fig. 9), this lead to an increase from the expected values of global SAR of up to 36% and peak SAR underestimated by up to 20%. In addition, the ratio of peak SAR to global SAR was underestimated by as much as 37%. Overall, the prediction error for SAR values ranged from about 19% to 40%. We also found that using a higher acceleration (shorter RF pulses) in Tx-SENSE, increased SAR error significantly.

## 5. CONCLUSIONS

In this study, realistic variations between simulated and constructed Tx-arrays were used to show that using improper coupling for determining simulated B1+ fields used in RF pulse calculations and corresponding Bloch simulations for predicting small FOV excitation leads to inaccuracies and therefore limits the ability to use simulations to optimize Tx-array designs for improved excitation profile. With the same model, it was shown that incorrect coupling for determining simulated E-Fields, leads to a global SAR prediction overestimated by up to 36% and a local SAR prediction underestimated by up to 39%.  Overall, the simulation-predicted peak/global SAR was underestimated between 19% and 40% in this model, generally worse for shorter (more accelerated) RF pulses.

Although, SAR values cannot be easily measured, proper coupling in simulations is absolutely necessary for RF coil designers to optimize the Tx-array design for Tx-SENSE and for accurate prediction of peak local SAR distributions. From the patient safety point of view, an underestimated local SAR prediction, normally scaled to the global SAR prediction, is particularly troublesome and may need to be incorporated into the safety plan for any site using a transmit-array, by setting more conservative limits for RF power.

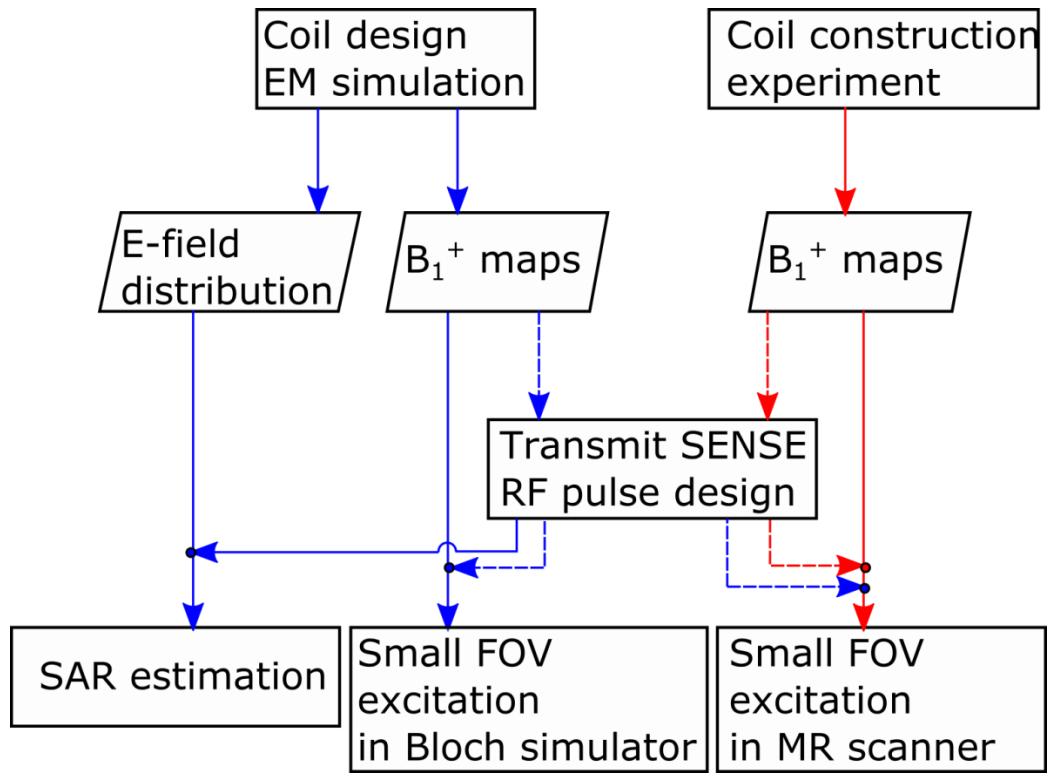

Fig. 1 Flowchart showing the coil optimization for small FOV excitation and SAR estimation in a simulation (blue) and in an experiment (red).

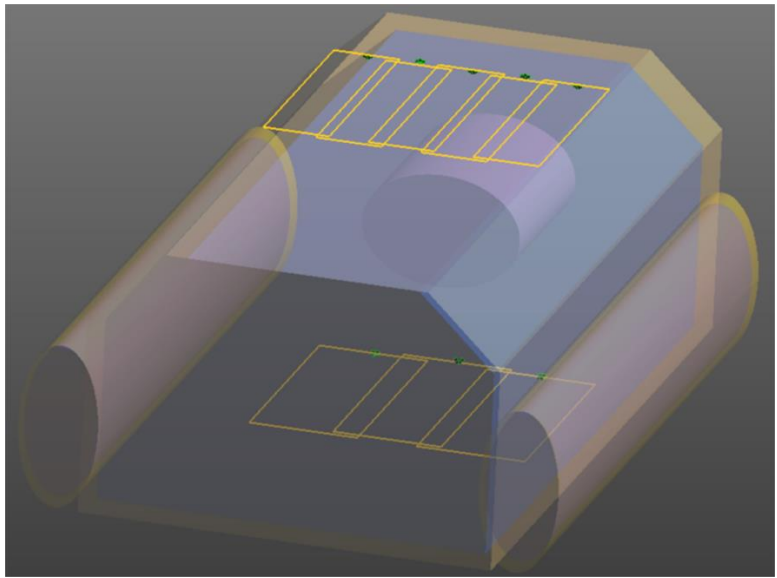

Fig. 2 Schematic showing the Tx-array coil configuration and the torso phantom for evaluating the impact of inductive coupling.

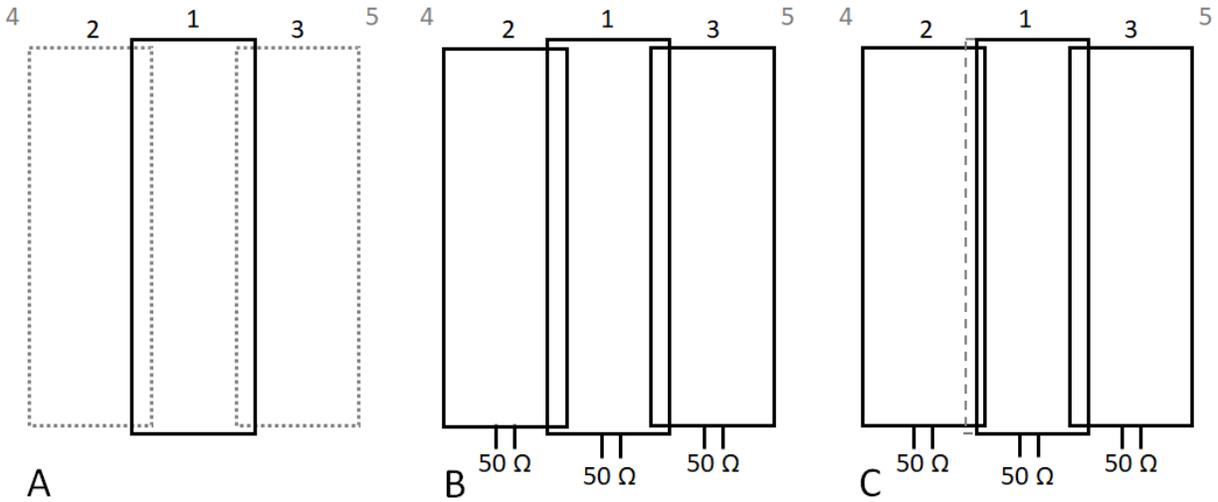

Fig. 3 Diagrams showing different scenarios of coupling for the model shown in Fig. 2, (A) isolated, (B) overlap decoupled, and (C) coupled coil elements. In (C) one conductor of the central element is shifted by 3mm from the dashed line to mimic a small variation during coil construction.

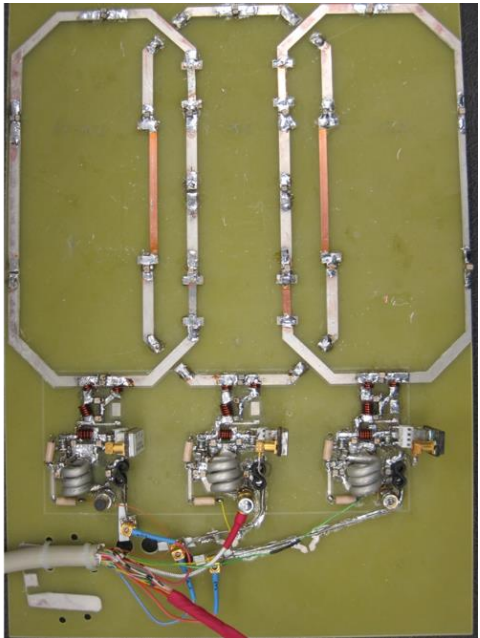

Fig. 4 Photograph showing the constructed coil corresponding to the posterior section of Fig. 2. Longitudinal conductors of the central element shown adjusted for zero mutual inductance with the neighboring elements, as in Fig. 3B.

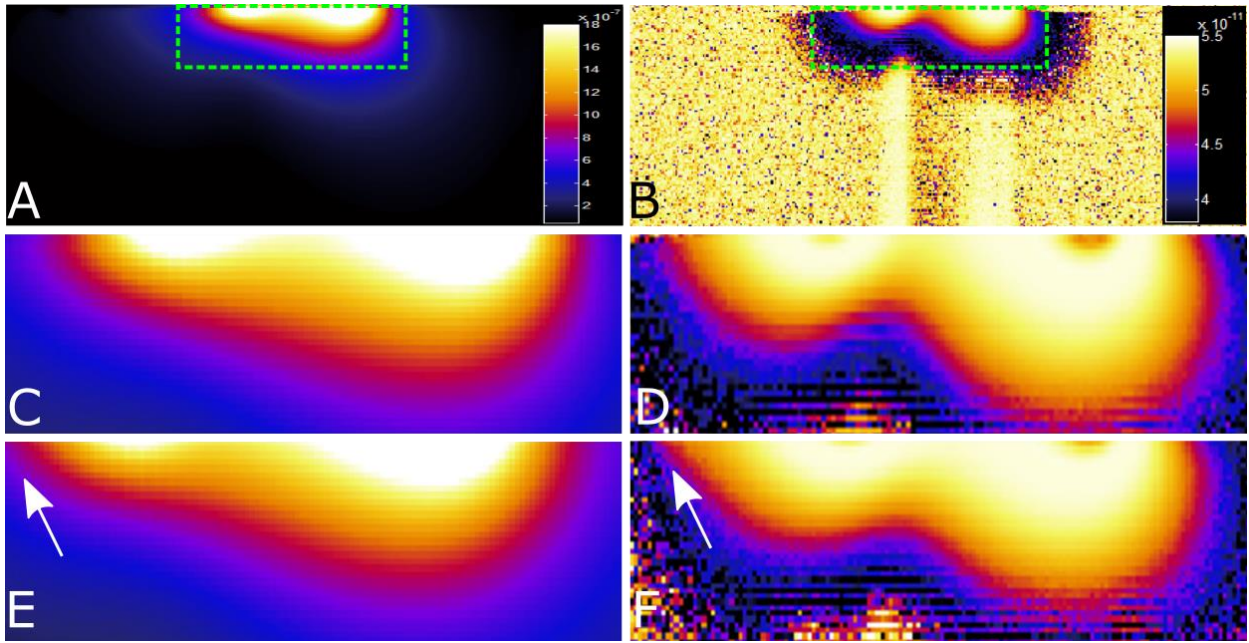

Fig. 5 Images showing B$_1^+$ maps of the central element where the left column is the simulated results and the right column is the experimental measured results for isolated (A through D) and coupled (E and F) scenarios. Images C and D are enlargements of the dashed regions shown in images A and B. The arrows show the regions where there is crosstalk with a neighboring element. Here three elements were located at the top of the images and only the central element was used to transmit and to receive.

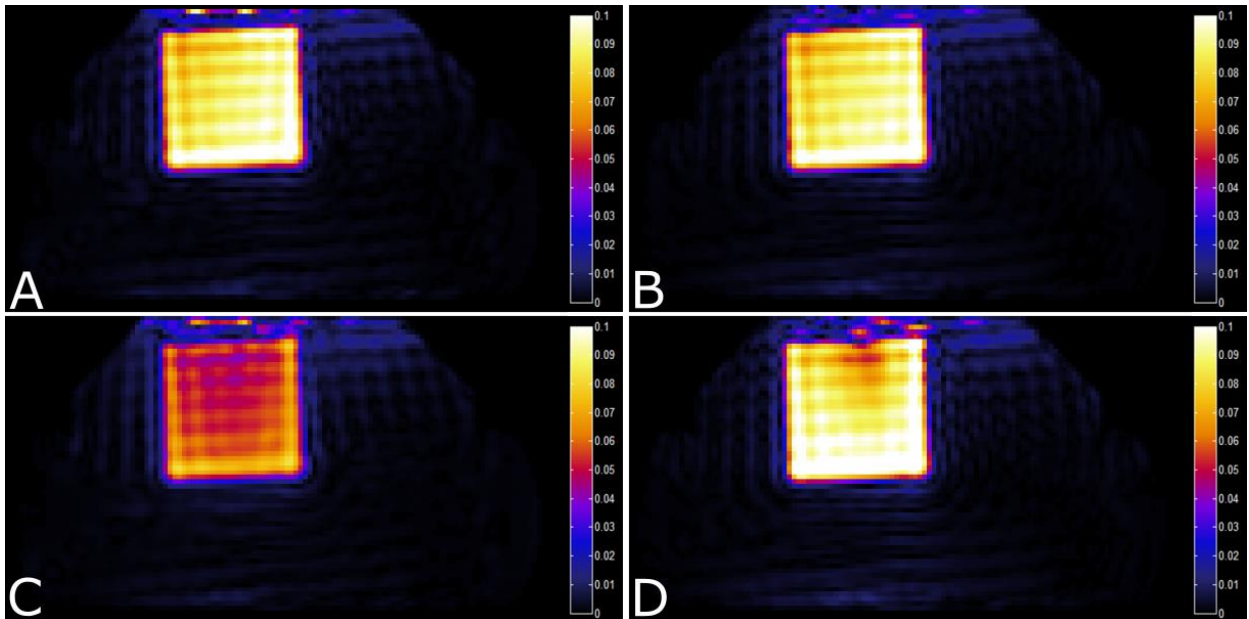

Fig. 6 Images showing simulated small FOV excitation when both RF pulses and B$_1^+$ maps were either in isolation (A) or overlap decoupled (B) scenarios. In (C) the RF pulses assumed no coupling but the coil used

overlap decoupling. In (D) the RF pulses assumed overlap decoupling, but the coil was not perfectly overlap decoupled. The color bar corresponds to the transverse magnetization relative to $M_0$.

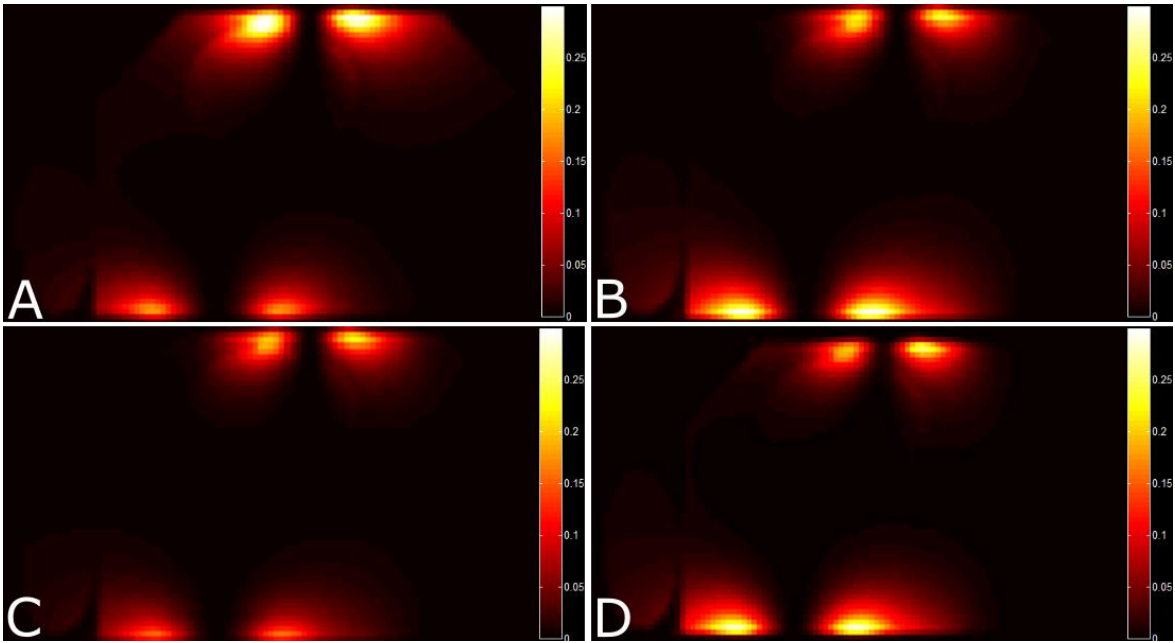

Fig. 7 Images showing simulated SAR distributions when RF pulses and E-field distributions were either in isolation (A) or were overlap decoupled (B). In (C) the RF pulses assumed no coupling but the coil used overlap decoupling. In (D) the RF pulses assumed overlap decoupling, but the coil was not completely overlap decoupled. The color bar corresponds to the SAR value per voxel. For comparison, the slice position of the SAR maps shown in Fig. 7A through D are identical to that of the small FOV excitation (Fig. 6).

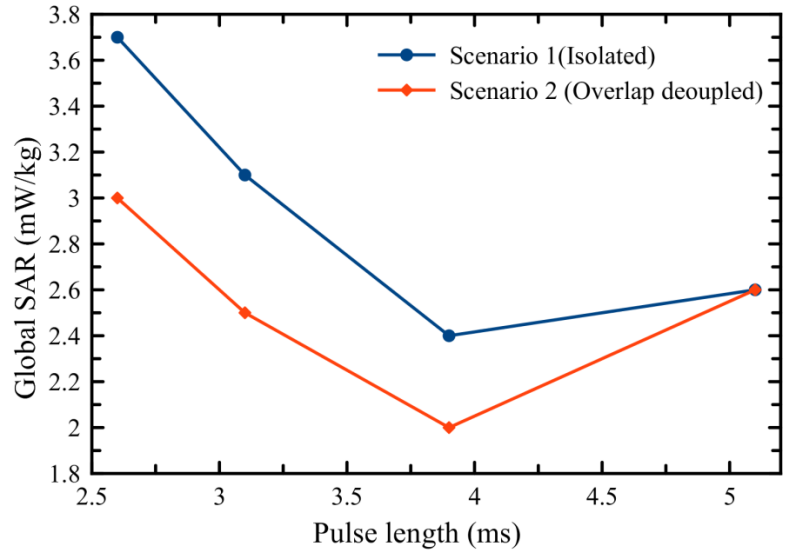

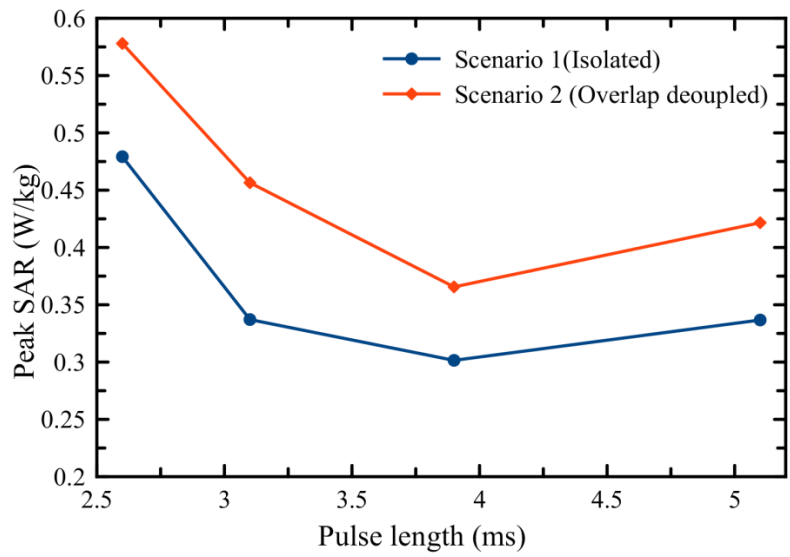

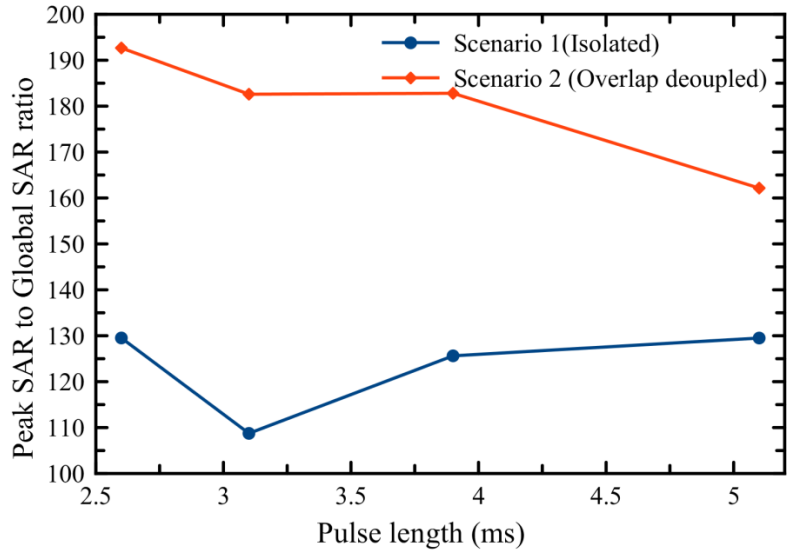

Fig. 8 Graphs showing global SAR, peak SAR, and global-to-peak SAR ratio as a function of RF pulse length determined from simulations of the coil shown in Fig. 3A and 3B. Scenario 1 (Blue Circles) use the E-field distribution and RF pulses determined with isolated elements. Scenario 2 (Red Diamonds) use the E-field distributions and RF pulses determined using overlap decoupled elements.

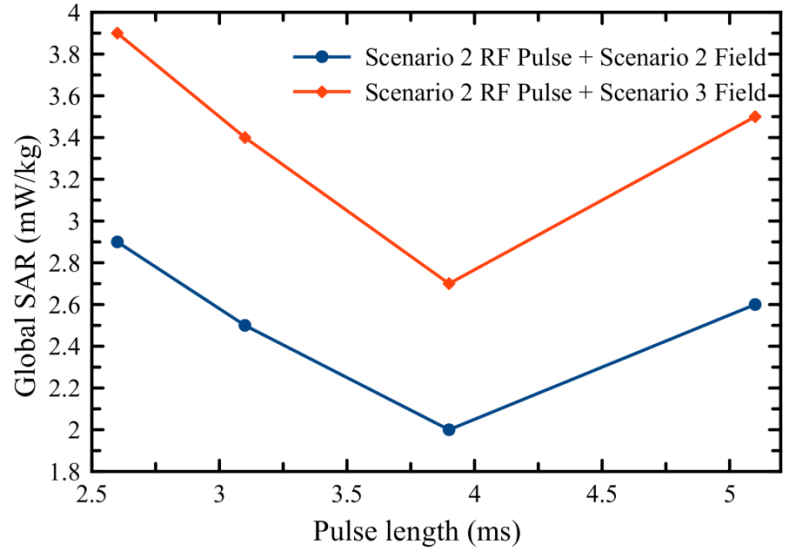

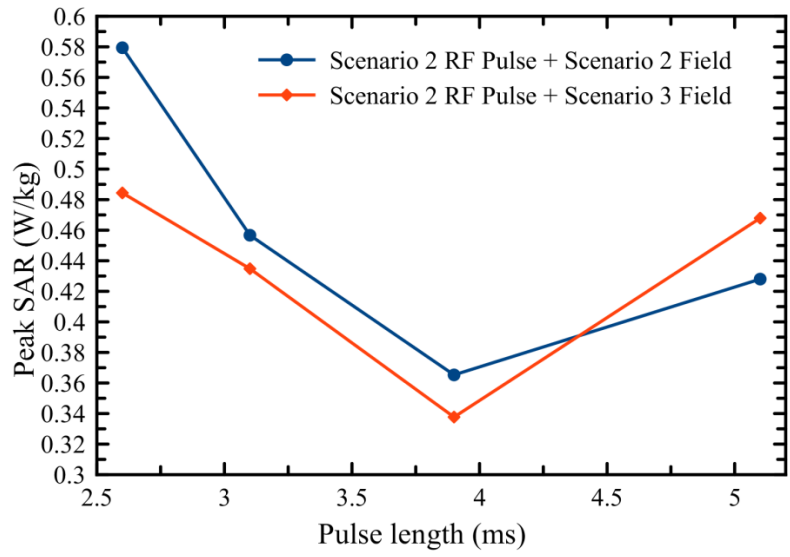

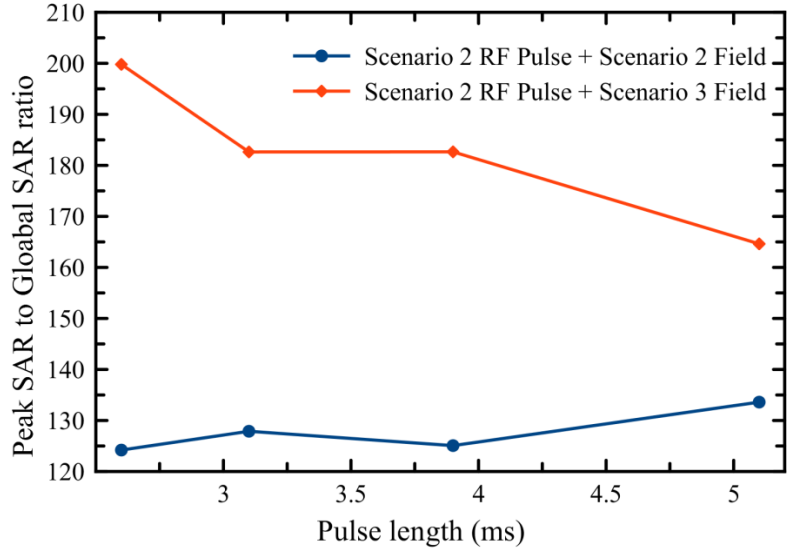

Fig. 9 Graphs showing global SAR, peak SAR, and peak-to-global SAR ratio as a function of RF pulse length determined from simulations of the coil shown in Fig. 3B and Fig. 3C. Blue Circles: RF pulses and E-field distributions determined from overlap decoupled elements illustrated in Fig. 3B. Red Diamonds: RF pulses determined from overlap decoupled elements illustrated in Fig. 3B but E-field distributions determined from coupled elements illustrated in Fig. 3C.

ACKNOWLEDGEMENTS

The authors acknowledge funding from the Canadian Breast Cancer Foundation and the National Sciences and Engineering Research Council of Canada.